\documentclass[11pt,subeqn]{article}

\usepackage{amssymb,amsfonts,amsmath}

\newcommand{\ket}{\rangle}
\newcommand{\bra}{\langle}

\newcommand{\cA}{{\mathcal A}}
\newcommand{\bA}{{\mathbf A}}
\newcommand{\bB}{{\mathbf B}}
\newcommand{\cB}{{\mathcal B}}
\newcommand{\bC}{{\mathbb C}}

\newcommand{\cF}{{\mathcal F}}

\newcommand{\cH}{{\mathcal H}}

\newcommand{\cL}{{\mathcal L}}

\newcommand{\bp}{{\bf p}}
\newcommand{\bP}{{\mathbb P}}
\newcommand{\bq}{{\bf q}}
\newcommand{\bR}{{\mathbb R}}
\newcommand{\res}{\text{res}}

\newcommand{\cT}{{\mathcal T}}

\newcommand{\sps}{\text{SPS}}
\newcommand{\non}{\nonumber}
\newcommand{\bvert}{{\Big \vert}}

\newcounter{mnotecount}[section]

\begin{document}
\begin{titlepage}
\vspace{.5in}
\flushright{hep-th/0508221\\
            August 2005}\\
\vspace*{.5cm}
\begin{center}
{\Large\bf  Symmetries of WDVV equations}\\[2ex]

Yujun Chen \footnote{ychen@perimeterinstitute.ca}\\
\vspace{.1cm}
       {\small\it Perimeter Institute}\\
       {\small\it 31 Caroline St. N., Waterloo, ON N2L 2Y5, Canada}\\[2ex]
Maxim Kontsevich \footnote{maxim@ihes.fr}\\
\vspace{.1cm}
       {\small\it IH\'{E}S, 35, route de Chartres}\\
       {\small\it F-91440 Bures-sur-Yvette, France}\\[2ex]
Albert Schwarz \footnote{schwarz@math.ucdavis.edu}\\
\vspace{.1cm}
       {\small\it Department of Mathematics}\\
       {\small\it University of California}\\
       {\small\it Davis, CA 95616, USA}
\end{center}
\vspace*{1cm}
\begin{center}
{\large\bf Abstract}
\end{center}
\begin{center}
\begin{minipage}{5in}
{\small
\noindent We say that a function $F(\tau)$ obeys
WDVV equations, if for
a given invertible symmetric matrix $\eta^{\alpha \beta}$ and all $\tau \in \cT
\subset \bR^n$, the expressions
$c^{~ \alpha}_{\beta~\gamma}(\tau) = \eta^{\alpha
\lambda} c_{\lambda \beta \gamma}(\tau) = \eta^{\alpha \lambda}
\partial_{\lambda} \partial_{\beta} \partial_{\gamma} F$ can be
considered as structure constants of commutative associative algebra;
the matrix $\eta_{\alpha \beta}$ inverse to $\eta^{\alpha \beta}$
determines an invariant scalar product on this algebra.  A function
$x^{\alpha}(z, \tau)$ obeying
$ \partial_{\alpha} \partial_{\beta} x^{\gamma} (z, \tau)
   = z^{-1} c^{~\varepsilon}_{\alpha ~ \beta} \partial_{\varepsilon}
     x^{\gamma} (z, \tau)
$
is called a
calibration of a solution of WDVV equations.  We show that there exists
an infinite-dimensional group acting on the space of calibrated
solutions of WDVV equations (in different form such a group was
constructed in [2]).  We describe the action of Lie algebra of this group.
\vspace*{.25cm}
}
\end{minipage}
\end{center}
\end{titlepage}

\addtocounter{footnote}{-1}

\section{Introduction}
\label{sec:indroduction}
\setcounter{equation}{0}

Two-dimensional topological quantum field theory can be identified
with a solution of WDVV equations, or, in geometric terms, with a
Fr{o}benius manifold.  In such a theory the algebra of observables
$\cH$ depends on parameters $\tau_1, \dots, \tau_n$ (on a point of a manifold $\cT$). Here
$n = \text{dim} \, \cH$.  Two-point correlation functions
\begin{align}
  \eta_{\alpha \beta} = \bra e_{\alpha} e_{\beta} \ket
\end{align}
determine nondegenerate bilinear inner product $(~ ,\, )$ on $\cH$.
Three-point correlation functions
\begin{align}
  c_{\alpha \beta \gamma}
  = \bra  e_{\alpha} e_{\beta}  e_{\gamma} \ket
\end{align}
determine structure constants
\begin{align}
  c^{~\alpha}_{\beta ~ \gamma}(\tau) = \eta^{\alpha \lambda}
                                 c_{\lambda \beta \gamma} (\tau)
\end{align}
of the algebra $\cH(\tau)$, which is commutative and associative
(in the above formulas $(e_1, \dots, e_n)$ is a basis of vector
space $\cH$ of observables; we assume that the vector space $\cH$
and the basis $(e_1, \dots, e_n)$ do not depend on $\tau$, but the
correlation function $\bra  e_{\alpha} e_{\beta}\,  e_{\gamma}
\ket$ depends on $\tau$).

One assumes that
$\eta_{\alpha \beta}$ does not depend on $\tau$ and the expression
$\partial_{\alpha} c_{\beta \gamma \lambda}$ is symmetric with respect
to indices $\alpha, \beta, \gamma, \lambda$ (we use the notation
$\partial_{\alpha}$ for the derivative $\partial/
\partial \tau^{\alpha}$).  This means that there exists a function
$F(\tau_1, \dots, \tau_n)$, the free energy, obeying
\begin{align}
\label{cf}
  c_{\alpha \beta \gamma}(\tau) = \partial_{\alpha \beta \gamma} F(\tau).
\end{align}
The function $F$ satisfies the WDVV equations
\begin{align}
\label{wdvv1}
  \sum_{\delta,\gamma=1}^n \frac{\partial^3 F(\tau) }{\partial \tau^{\alpha}
   \partial \tau^{\beta} \partial \tau^{\delta} } \eta^{\delta \gamma}
   \frac{\partial^3 F(\tau)}{\partial \tau^{\gamma} \partial \tau^{\omega}
   \partial \tau^{\rho}}
   =\sum_{\delta, \gamma=1}^n
  \frac{\partial^3 F(\tau)}{\partial \tau^{\alpha} \partial
   \tau^{\omega} \partial \tau^{\delta} }
  \eta^{\delta \gamma} \frac{\partial^3 F (\tau) }{\partial \tau^\gamma
   \partial \tau^\beta \partial \tau^\rho },
\end{align}
reflecting associativity of the algebra with structure constants
$ c^{~\alpha}_{\beta ~ \gamma}(\tau) = \eta^{\alpha \lambda}
c_{\lambda \beta \gamma} (\tau)$. Usually one requires that the first
element $e_1$ of the basis of $\cH$ is the unit element and
$\eta_{\alpha \beta} = c_{1 \alpha \beta}$, then
\begin{align}
\label{wdvv2}
  \frac{\partial^3 F}{\partial \tau^\alpha \partial \tau^\beta \partial \tau^1 }
  = \eta_{\alpha\beta}.
\end{align}
It would be convenient
for us to impose the condition (\ref{wdvv2}) only at the end of our
consideration.

In the above formula we assumed that the vector space $\cH$
of observables does not depend on the parameters $\tau_1, \dots, \tau_n$ .  This assumption is justified by a remark  that starting with one of the points of the
manifold $\cT$ one can obtain all other points of this manifold using perturbation theory (every observable  can be considered as a perturbation). This construction can be used to specify a natural identification of all tangent spaces to this manifold . These tangent spaces can be considered
as Frobenius algebras (associative commutative algebras with
non-degenerate inner product).  The geometry arising on $\cT$ was
analyzed in \cite{dubrovin}.  Using the terminology of \cite{dubrovin}
one can say that $\cT$ is
equipped with a structure of Frobenius manifold.  

Let us define
operators $\nabla_{\alpha}(z)$ on the space of $\cH$-valued functions
on $\cT$ by the formula
\begin{align}
  \nabla_{\alpha} = \partial_{\alpha} - z^{-1} \hat{c}_{\alpha}.
\end{align}
Here $z$ is a complex parameter; if $\varphi = \varphi^{\gamma}
e_{\gamma}$, then
$\hat{c}_{\alpha} \varphi = c^{~\beta}_{\alpha ~ \gamma}
\varphi^{\gamma} e_{\beta}$.  It is easy to check that operators
$\nabla_{\alpha}$ corresponding to $c_{\alpha \beta \gamma}$,
$\eta_{\alpha \beta}$ obeying WDVV equations satisfy
\begin{subequations}
\begin{align}
\label{nabla1}
  &[\nabla_{\alpha}(z),\nabla_{\beta}(z) ]= 0, \\
\label{nabla2}
  &(\nabla_{\alpha}(z) \, \varphi, \psi) + (\varphi,
  \nabla_{\alpha}(-z)\,  \psi)=\partial_{\alpha}
(\varphi, \psi).
\end{align}
\end{subequations}
Conversely, if (\ref{nabla1}) and (\ref{nabla2}) are satisfied, then
$c_{\alpha \beta \gamma}$ and $\eta_{\alpha \beta}$ correspond to a
solution of WDVV equations.

It follows from (\ref{nabla1}) that there exists a function
$S(\tau, z)$ defined for $z\neq 0$, which  takes values in invertible $n \times
n$ matrices (or, speaking
in more invariant way, in the group Aut$\, \cH$ of automorphisms of
vector space $\cH$) and obeys
\begin{align}
  \nabla_{\alpha}(z) S(\tau,z)=0,
\end{align}
or, more precisely
\begin{align}
\label{S_equation}
  \partial_{\alpha} S^{\beta}_{~\gamma} = z^{-1}
c^{~\epsilon}_{\alpha ~ \gamma}
  S^{\beta}_{~\epsilon}.
\end{align}
Using (\ref{nabla2}) one verifies that $S(z)$ can be chosen (non-uniquely) in
such a way that the normalization conditions
\begin{subequations}
\begin{align}
\label{S_normalization_a}
  S(z=\infty) &=1\\
\label{S_normalization_b}
  S(\tau,z) S^*(\tau,-z)&=1.
\end{align}
\end{subequations}
are satisfied.  Here $S^*$ stands for an operator adjoint to $S$
with respect to bilinear inner product $(~,\,)$.
The choice of $S(\tau,z)$ is called calibration of Frobenius manifold (of
solution of WDVV equations).

We assume that $S(z)$ is a holomorphic function of $z
\in \bP^1\backslash \{ 0 \}$ (here $\bP^1 = \bC \cup \{\infty \}$) .  It is clear
that two solutions of (\ref{S_equation}) obeying normalization
conditions (\ref{S_normalization_a}) and (\ref{S_normalization_b}) are
related by the formula
\begin{align}
  \tilde{S}(\tau,z) = S(\tau,z) M(z),
\end{align}
where $M$ is a holomorphic function on $\bP^1 \backslash \{ 0 \}$
obeying
\begin{subequations}
\begin{align}
\label{M_1}
  M(z=\infty)&=1,\\
\label{M_2}
  M(z) M^*(-z)&=1.
\end{align}
\end{subequations}
In other words, the choice of calibration is governed by the group
$\bB_+$ consisting of holomorphic matrix functions on $\bP^1\backslash
\{ 0 \}$ satisfying
(\ref{M_1}) and (\ref{M_2}).  Notice that the functions
$S^{\alpha}_{~\beta}(\tau, z)$ obey
\begin{align}
\label{symmetry}
  \partial_{\lambda} S^{\alpha}_{~\beta} = \partial_{\beta}
    S^{\alpha}_{~ \lambda},
\end{align}
which follows from symmetry of structure constants:
$c^{~\epsilon}_{\beta~\lambda}= c^{~\epsilon}_{\lambda~\beta}$.
Using (\ref{symmetry}) one can construct a function
$x^{\alpha}(\tau,z)$ satisfying
\begin{align}
  S^{\alpha}_{~\beta}(\tau,z) = \partial_{\beta} x^{\alpha} (\tau,z);
\end{align}
this function is defined up to a holomorphic summand that does not
depend on $\tau$ (i.e. we can take $x^{\alpha}(\tau,z) +
\rho^{\alpha}(z)$ instead of $x^{\alpha}$).  For fixed $z$ we can
consider $(x^1(\tau, z), \dots, x^n(\tau,z))$ as coordinates on
manifold $\cT$; these coordinates are defined up to affine
transformations (a choice of basis in the space of observables and a
choice of calibration result in linear transformations of these
coordinates, the freedom in a choice of $x^{\alpha}$ for given $S$
leads to a shift).

  In what follows we will modify the notion of calibration, saying
that a calibrated solution of WDVV equations is a solution of these
equations together with a solution of equations
\begin{align}
\label{xequation}
  \partial_{\beta} \partial_{\gamma} x^{\alpha} = z^{-1} c^{~\,
  \epsilon}_{\beta ~ \gamma} \partial_{\epsilon} x^{\alpha},
  ~~~~\alpha =1,2
\end{align}
obeying normalization conditions
\begin{align}
  x^{\alpha}(\tau,\infty)=\tau ^\alpha,\\
\label{normalization1}
   \eta  _{\alpha \beta} \partial_{\lambda} x^{\alpha}(\tau,z) \partial_{\mu}
  x^{\beta}(\tau, -z) &= \eta _{\lambda \mu}.
\end{align}
Let us consider holomorphic functions on $\bC^{\times} = \bC
\backslash \{0\}$, which take values in the group $GL(\cH)$ of
invertible
linear transformations of $\cH$ or, more generally, in the group
$Af\!f(\cH)$ of invertible affine transformations of $\cH$.  Using
bilinear inner product on $\cH$ we define a group $\bB$ (twisted loop
group) as a group of $GL(\cH)$-valued holomorphic functions $B(z)$
on $\bC^{\times}$ obeying
\begin{align}
\label{B_norm}
   B(z) B^*(-z) =1.
\end{align}
Similarly, we define a group $\bA$ as a group of $Af\!f(\cH)$-valued
functions $(B(z), d(z))$, where the linear part $B(z)$ obeys
(\ref{B_norm}). Here we write an affine transformation as a pair $(B,d)$,
where $B$ is a linear transformation and $d$ is a shift:
\begin{align}
\label{affine1}
  x \rightarrow B x + d.
\end{align}
Lie algebras of groups $\bB$ and $\bA$ are denoted by $\cB$ and
$\cA$ correspondingly. Notice that in the modified definition a
choice of calibration is governed by the subgroup ${\bf A_+}$ of
the group ${\bf A}$ consisting of transformations of the form
(1.20) with $B\in {\bf B_+}$.

Let us take a calibrated solution
$(F(\tau), x^{\alpha}(\tau, z))$ of WDVV equations and an element
 $(b(z), d(z)) \in \cA$.  We define
\begin{align}
\label{fvari}
  \delta F (\tau)
  =&  \,- \frac{1}{2 \pi i} \int_{\Gamma} \Big [\,
     \frac{b_{\alpha \beta}(\zeta) \,x^{\alpha} (\tau, \zeta) }{2} + d_{\beta}(\zeta)\, \Big
     ]\, x^{\beta}(\tau, -\zeta)  \, d\zeta, \\
\label{xvari}
  \delta x^{\alpha} (\tau, z)
 = & \, \frac{1}{2 \pi i} \int_{\Gamma}
       \frac{\eta^{\rho \epsilon} [\, b_{\lambda \sigma}(\zeta)
x^{\lambda}(\tau, \zeta)
     + d_{\sigma}(\zeta) ]\, \partial_{\epsilon} x^{\sigma} (\tau, -\zeta)
      }{\zeta-z} \,
       \partial_{\rho} x^{\alpha} (\tau , z)
       \, d \zeta.
\end{align}
Here $\Gamma$ is a circle with the center at $z=0$; in
(\ref{xvari}) we assume that the radius of this circle is less
than $\vert z\vert$. {\it We will prove that} $(F+ \delta F, x+ \delta x)$
{\it is a calibrated solution of} WDVV {\it equations.}  In other
words, the Lie algebra $\cA$ (or, more precisely, its extension )
acts on the space of calibrated solutions of WDVV equations.

We did not include the existence of unit element in our axioms of
two-dimentional TQFT.  If the existence of unit
is required, we consider transformations given by elements $(b(z),
d(z)) \in \cA$, where $d(z)$
is related to $b(z)$  in the following way
\begin{align}
\label{dandb}
  d_{\alpha}(z) =  z b_{1 \alpha}(z).
\end{align}
It follows from (1.6) that we can impose a  normalization condition for $x^{\alpha}
(\tau, z)$
\begin{align}
  \label{normalization2}
  \partial_1 x^{\alpha} (\tau, z) &= z^{-1} x^{\alpha}(\tau, z) + \delta_{~1}^
  {\alpha}.
\end{align}
Conversely, differentiating (\ref {normalization2}) and comparing with  (1.16) we obtain that (1.24) implies $e_1=1$.
The relation (1.23) is compatible with the
normalization condition (1.24) . More precisely, if we define the variation of
$x^{\alpha}$ by means of (1.22)  where $d(z)$ is related to $b(z)$
as in (1.23) the new $x^{\alpha}$ again obeys the same normalization
condition; hence $e_1$ is the unit element also after variation.

In this way we obtain an action of Lie algebra $\cB$ on the space of
calibrated Frobenius manifolds with unit elements.  
This action
corresponds to the action of twisted loop group constructed by Givental in \cite{givental1} . Notice that Givental's construction
provides an action of twisted loop group on the
space of genus 0 TQFT coupled to gravity. If the Frobenius manifold is constructed by means of  semi-infinite variation of Hodge structures the existence of the action of the twisted loop group
follows from results of Barannikov \cite{bar1}. In semisimple
case a different construction of the action of the same group was given by
van de Leur; see \cite{leur}).
Givental presented strong evidence that for all genera the same group acts on the
space of TQFT's coupled to gravity, at least at the level of perturbation
theory.  Recently one of us (M.K., in preparation) defined  the action of twisted loop group  also for open-closed theory confirming a conjecture of another of us (A. Sch.).

\section{Geometry of Frobenius manifolds}
\label{geometry}
\setcounter{equation}{0}

We have seen that coordinates
$x^{\alpha}(\tau, z)$ are defined up to affine transformations.  This
means that we obtained a family of affine structures $T_z$ on
manifold $\cT$ depending on parameter $z \in \bP^1\backslash \{ 0 \}$.
We can consider
the direct product of $\bP^1 \times \cT$ as a holomorphic bundle over $\bP^1$ ;
all fibers of this bundle except the fiber over $z=0$ are affine
spaces.

It is useful to give an invariant description of the above
structure.  To do this we notice that our construction of affine
structures $T_z$ can be regarded as a particular case of general
construction of affine structure by means of torsion-free flat
connection $\Gamma^{\gamma}_{\alpha \beta}$ (the connection is
torsion-free if $\Gamma^{\gamma}_{\alpha \beta}= \Gamma^{
\gamma}_{\beta \alpha}$;  it is flat if covariant derivatives
constructed by means of Christoffel symbols
$\Gamma^{\gamma}_{\alpha \beta}$ commute).  We can say that
$z^{-1} c^{~ \beta}_{\alpha~ \gamma}$ determines a family of
torsion-free flat connections on $\cT$ that has a pole of order
$1$ at $z = 0$. Conversely, let us consider a holomorphic bundle
$\cal E$ over $\bP^1$ with the fiber over $0 \in \bP^1$ identified
with $\cT$. Let us assume that all fibers over points $z \in \bP^1
\backslash \{0 \}$ are equipped with affine structure and that the
affine structure is defined by means of torsion-free flat
connection $\Gamma^{\gamma}_{\alpha \beta}(z)$  having a pole of
order $1$ at $z= 0$.  If the bundle at hand is holomorphically
trivial we can identify its total space with $\bP^1 \times \cT$;
this identification gives a coordinate system on the total space
where $\Gamma^{\gamma}_{\alpha \beta}$ is linear with respect to
$1/z$ (we use the fact that every holomorphic function of $z \in
\bP^1 \backslash \{0\}$ having first order pole at $z=0$ is linear
with respect to $1/z$).

  Due to
(\ref{normalization1}) one
can construct a nondegenerate bilinear pairing between tangent spaces
to affine spaces $T_z$ and $T_{-z}$ where  $z \in
\bP^1 \backslash \{0\}$ (notice that tangent spaces at
different points of affine space are identified). More precisely,
we can rewrite (\ref{normalization1}) as
$$ \eta _{\alpha \beta} \delta x^{\alpha}(\tau, z) \delta x^{\beta}(\tau, -z)=
\eta _{\mu  \nu} \delta \tau ^{\mu}\delta \tau ^{\nu} .$$
This equation shows that the metric $\eta _{\alpha \beta}$ induces covariantly constant pairing
between tangent spaces
to affine spaces $T_z$ and $T_{-z}$ ;  for fixed $\tau$ and $z$ tending
to zero this pairing has a limit ( in our coordinate system it does not depend on $z$) .
The statement about existence of limit remains correct if $\tau$ is not fixed, but depends analytically on $z$ in a neighborhood
of $z=0$. To analyze the case when $\tau = \tau (z)$ we
should prove that the expression
\begin{align}
  \label{pairing}
\eta _{\alpha \beta}\frac {\partial x^{\alpha}} {\partial \tau ^{\lambda}}\bvert_{(\tau = \tau (z),z)}\frac {\partial x^{\beta}} {\partial \tau ^{\mu}}\bvert_{(\tau = \tau (-z),-z)}
\end{align}
has a finite limit as $z$ tends to zero. To
use  the relation (\ref{normalization1})
we decompose
$$\frac {\partial x^{\alpha}} {\partial \tau ^{\lambda}}\bvert_{(\tau ,z)}$$ into Taylor series at the point $( \tau _0,z)$; we apply this decomposition to the case $\tau=\tau (z), \tau _0= \tau (-z)$.
One can check that for $z$ tending to zero the leading term
can be written in the form
\begin{align}
  \label{taylor}
\frac {\partial x^{\alpha}} {\partial \tau ^{\lambda}}\bvert_{(\tau ,z)}=\frac {\partial x^{\alpha}} {\partial \tau ^{\lambda}}\bvert_{(\tau _0 ,z)} +\sum_{n>0}\frac{1}{n!}z^{-n} ({\hat C}^n)_{\lambda}^{\sigma}\frac {\partial x^{\alpha}}{\partial \tau ^{\sigma}} .
\end{align}
Here $\hat C$ stands for the matrix $$C_{\rho}^{\nu}=
(\tau ^{\mu}-\tau _0^{\mu})c_{\rho \mu }^{\nu}(\tau_0).$$
Applying (\ref{taylor}) and (\ref{normalization1}) we obtain
that (\ref {pairing}) has a limit as $z$ tends to zero; it
converges to
\begin{align}
  \label{eta}
({{\rm exp}{\tilde C}})_{\lambda}^{\sigma}\eta _{\sigma \mu}
\end{align}
where $${\tilde C}_{\rho }^{\nu}=2\frac {d\tau ^\mu}{dz}\bvert _{z=0}c_{\rho \mu }^{\nu}(\tau_0).$$

In the derivation of (\ref {taylor}) one should use the formula
\begin{align}
  \label{a}
  \frac {\partial ^2 x^{\alpha}} {\partial \tau ^{\beta}\partial \tau^
  {\gamma}}=z ^{-1}c^{~\epsilon}_{\beta \, \gamma}\frac {\partial x^{\alpha}} {\partial \tau ^{\epsilon}}
  \end{align}
and similar formula expressing higher partial derivatives of
$x^\alpha$ in terms of the first derivatives (One obtains such formulas differentiating (\ref{a})).  We
assume that $z$ tends to zero and keep only the leading term;
then there is no necessity to take derivatives of $c^{\epsilon}_{\beta \gamma}$ differentiating (\ref{a}). We obtain that in our approximation
\begin{align}
 \label{b}
\frac {\partial ^{k+1} x^{\alpha}}{\partial \tau ^{\beta}\partial \tau ^{\gamma _1} \dots \partial \tau^ {\gamma _k}}=z^{-k}c_{\gamma_k ~\rho _{k-1}}^{~\rho _k}c_{\gamma_{k-1} ~\rho _{k-2}}^{~\rho _{k-1} }...c_{\gamma_1 ~\beta}^{~\rho _1}\frac {\partial x^{\alpha}} {\partial \tau ^{\rho _1}}
\end{align}
It is easy to check that (\ref {taylor}) follows from (\ref {b}).

 Let us describe now geometric
data that permit us to construct a calibrated solution of WDVV equations. We
start with holomorphic bundle over $\bP^1$ and a family of
torsion-free flat connections $\Gamma^{\gamma}_{\alpha
\beta}(z,\tau)$ on the fibers $T_z$ of this bundle that depend
holomorphically on $z \in \bP^1 \backslash \{0\} $ and have a pole
of order $1$ at $z=0$.  We assume that the bundle is
holomorphically trivial;  trivialization permits us to represent
the connections in the form $\Gamma^{\gamma}_{\alpha
\beta}(z,\tau)= z^{-1} c^{~ \gamma}_{\alpha ~ \beta}(\tau)$ in
corresponding coordinate system.

 Let us suppose in
addition that we have
a covariantly constant nondegenerate bilinear pairing between tangent spaces to fibers over $z$ and over
$-z$ for $z\not= 0$ that depends holomorphically on $z$.(As we noticed affine structure on
$T_z$ permits us to identify
tangent spaces at all points of $T_z$). This pairing gives a flat metric on $T_{\infty}$.  We assume that  the pairing between tangent spaces to $T_z$ and $T_{-z}$  has a limit  if the points of $T_z$ and
$T_{-z}$ tend to a point of $T_0$ staying on a holomorphic  section of our bundle over
a neighborhood of $z=0$. (It is sufficient to assume that the pairing is bounded, because a  holomorphic  function bounded in a
neighborhood of a point has a removable singularity at this point.)

Writing the bilinear pairing in the coordinate system
coming from a trivialization of the bundle we
obtain that
the pairing can be described by means of matrix $\eta_{\alpha \beta} (\tau)$
that does not depend on $z$. (Independence of $z$
follows
from the fact that a bounded holomorphic function of $z$ is a
constant.) We conclude from our assumption about the fiber $T_{\infty}$ that the matrix $\eta_{\alpha \beta} (\tau)$ specifies a nondegenerate flat metric on fibers.

 It follows from above remarks that  $c^{~
\gamma}_{\alpha ~ \beta}$ and $\eta_{\alpha \beta}$ obey (1.8a) and (1.8b), hence they specify a
solution to WDVV equation; the choice of trivialization of affine bundle over  $ \bP^1 \backslash \{0\}$corresponds to
the choice of calibration.

\section{Symmetries of WDVV equation}
\label{Kontsevish}
\setcounter{equation}{0}

Let us consider a solution to WDVV equations and its calibration
$x^{\alpha}(\tau, z)$.  As we have seen, the functions
$x^{\alpha}(\tau,z)$ determine an affine structure on fibers
of the trivial bundle over $\bP^1$
except the fiber over $z=0$.  One can obtain $\bP^1$ pasting
together two open disks $|z| < \infty$ and $|z| > 0$.  We can
construct a new holomorphic bundle over $\bP^1$ twisting the
direct product $\bP^1 \times \cT$ over $\bC^{\times}$ , which is
the common region of these two disks $D_0$ and $D_{\infty}$ .  The
total space of the new bundle is obtained by means of
identification of $\bC^{\times} \times \cT \subset D_0 \times \cT$
with $\bC^{\times} \times \cT \subset D_{\infty} \times \cT$ by
the formula $(z, \tau) \sim (z, f_z(\tau))$. Here
$f_z: T_z \rightarrow T_z$ is an analytic
automorphism of $T_z$ (of the fiber over $z \in
\bC^{\times}$) that depends analytically of $z \in
\bC^{\times}$.
  The fiber $T_z$ is equipped with affine
structure.  We will assume that $f_z: T_z \rightarrow
T_z$ is an affine transformation;  then the fibers of the new
bundle over all points $z \in \bP^1 \backslash \{0\}$ also can be
considered as affine spaces.

If $f_z$ is sufficiently close to identity the new holomorphic
bundle is holomorphically trivial (this follows from the remark that
holomorphically trivial vector bundles form an open subset of the
space of holomorphic vector bundles over $\bP^1$).  As we
know, there
exists a  pairing between tangent spaces to the
fibers $T_z$ and $T_{-z}$.  To guarantee the existence
of similar pairing for the new bundle we
should impose some conditions on affine maps
$f_z: T_z \rightarrow T_z$.  Namely, if
\begin{align}
\label{affine}
  f_z = (B(z ), d(z))
\end{align}
we should require
\begin{align}
\label{require}
  B(z) B(-z)^*=1.
\end{align}
Notice that we did not change the pairing  over the inner disk,
therefore the behavior of the pairing as $z$ tends to zero does not
change. We imposed the condition that the pairing is bounded
on every holomorphic section over a neighborhood of
$z=0$ ; this condition is fulfilled for the new bundle. Hence
the new bundle also specifies a calibrated solution to WDVV equations.

The group of matrix functions $B(z)$ obeying (\ref{require})
is a (version of) twisted loop group of \cite{givental1}.  The group of affine
transformations of the form (\ref{affine}) with the linear part
obeying (\ref{require}) was denoted by $\bA$ in Section
\ref{sec:indroduction} and the corresponding Lie
algebra was denoted by $\cA$. In what follows it will be convenient to consider
$f_z$ as an operator on the vector space of observables $\cH$.

It follows from the above consideration that the Lie algebra $\cA$
acts on the space of calibrated solutions of WDVV equations.  Our goal
is to calculate this action more explicitly.

Elements of $\cA$ can be considered as pairs $(b(z),
d(z))$, where $b(z)$ is a holomorphic function on $\bC^{\times}$
with values in linear maps $\cH \rightarrow \cH$ obeying $b(z) +
b(-z)^* =0$ and $d(z)$ is a holomorphic $\cH$-valued
function on $\bC^{\times}$.
  To calculate the action of $(b(z),
d(z))$ we should trivialize the twisted bundle.  This means that
for every $\tau \in \cT = T_{\infty}$, we should find a holomorphic
section of the twisted bundle taking the value $\tau$ at $z=\infty$.  We
will write this holomorphic section as a pair of holomorphic sections
of direct product $D_0 \times \cT$ and $D_{\infty} \times \cT$ with
appropriate
gluing conditions on $\bC^{\times}$.   Representing sections of $D_i
\times
\cT$ as $\cT$-valued functions we can write down the gluing condition in
the form
\begin{align}
\label{twist}
   (\delta^{\mu}_{~\lambda} + b_{\lambda}^{~\mu}(z))
 x^{\lambda}
    (\kappa(z), z) + d^{\mu}(z)
    = x^{\mu} (\alpha (z),z),
\end{align}
where $\alpha^{\rho}(z)$ and $\kappa^{\rho}(z)$ close to $\tau^{\rho}$
can be expressed as
\begin{align}
  \alpha^{\rho}(z) &= \tau^{\rho} + a^{\rho} (z),
  ~~~~~~\text{inner disk}~ D_0 ~,
  \non \\
  \kappa^{\rho}(z)  &= \tau^{\rho} + k^{\rho}(z),
   ~~~~~~ \text{outer disk} ~D_{\infty} ~.
\end{align}
Here we are working with infinitesimal group transformation, corresponding
to Lie algebra element $(b(z), d(z))$. We require
$k(\tau, \infty) =0$ to get a section containing
$\tau \in \cT= T_{\infty}$.

From (\ref{twist}) we have
\begin{align}
\label{bc}
  \varphi^{\rho}(z) \equiv a^{\rho}(z) - k^{\rho}(z)
 = (S^{-1}(z))^{\rho}_{~\mu}\,
   [\, b^{~\mu}_{\lambda}(z) x^{\lambda}(z)
      +d^{\mu} (z) \, ].
\end{align}
Knowing (\ref{bc}), we can express $a$ and $k$ in terms of Cauchy integral
\begin{align}
\label{cauchy}
  \Phi^{\rho}(z) = \frac{1}{2 \pi i} \int_{\Gamma}
\frac{\varphi^{\rho}(\zeta)}{\zeta-z} \, d \zeta,
\end{align}
where $\Gamma$ is any circle with the center at the origin. This
integral represents a function that is holomorphic everywhere
except for $\Gamma$, on which it has a jump $\varphi^{\rho}$, and
then tends to zero at infinity. Using these facts we obtain
\begin{align}
\label{holo}
  a^{\rho}(z) = \Phi^{\rho}(z)
\end{align}
inside $\Gamma$,
\begin{align}
\label{hol}
  k^{\rho} (z) = \Phi^{\rho}(z)
\end{align}
outside $\Gamma$ and
\begin{align}
  k^{\rho}(z) \rightarrow 0 \text{~~~~~as~} z \rightarrow \infty .
\end{align}
The expression (\ref{hol}) for $k^{\rho}$  leads
immediately to the expression (\ref{xvari}) for
$\delta x^{\alpha}$.

To calculate the variation of structure constants $c$ we notice that
we do not change the connection over the inner disk; the only change in $c$
comes from the change of coordinates. Using the standard rules for the
variation of connection one can obtain that the structure constants
behave like a tensor
by the change of coordinates; this fact was used in [3]. (Recall, that structure constants
describe the behavior of connection at the point $z=0$ where the connection
has a pole .) The change of coordinate $\tau$
  at $z=0$ is governed by the formula
\begin{align}
\label{var_tau}
   a^{\rho}(z=0) = \frac{1}{2 \pi i} \int_{\Gamma}
\frac{\varphi^{\rho} (\zeta)}{\zeta} \, d \zeta=
    \frac{1}{2 \pi i} \int_{\Gamma}
        \frac{\eta^{\rho \epsilon} [b_{\lambda \sigma}(\zeta)
        x^{\lambda}(\zeta) + d_{\sigma}(\zeta)] \,
         \partial_{\epsilon} x^{\sigma}(-\zeta)}{\zeta} \, d \zeta.
\end{align}

Calculating the Jacobian matrix at $z=0$
\begin{align}
  L^{\mu}_{~\nu} = \frac{\partial \tau^{\mu}}{\partial
                    \alpha^{\nu}}
                 = \delta^{\mu}_{~\nu} &- \frac{1}{2 \pi i}
                 \int_{\Gamma} \frac{\eta^{\mu \epsilon} b_{\lambda
                 \sigma}(\zeta)}{\zeta}\, \big [\partial_{\nu} x^{\lambda}
                 (\zeta) \partial_{\epsilon} x^{\sigma} (-\zeta)
                  + x^{\lambda} (\zeta)
                  \partial_{\nu} \partial_{\epsilon} x^{\sigma}
                 (-\zeta) \big ]d \zeta \non \\
                & - \frac{1}{2 \pi i}
                   \int_{\Gamma} \frac{\eta^{\mu \epsilon}\,
                   d_{\sigma}(\zeta)}{\zeta}\,
                   \partial_{\nu} \partial_{\epsilon}
                    x^{\sigma} (-\zeta)\, d \zeta,
\end{align}
we obtain the variation of structure constants
\begin{align}
\label{cvari}
  \delta c^{~\beta}_{\alpha~\gamma} (\tau)
  =&  (L^{-1})^{\mu}_{~\alpha} \, L^{\beta}_{~\nu} \,
      (L^{-1})^{\xi}_{~\gamma} \,
      c^{~\nu}_{\mu~\xi} (\tau) - c^{~\beta}_{\alpha~\gamma} (\tau)
      - \partial_{\epsilon} (c^{~\beta}_{\alpha~\gamma} )\, \delta
      \tau^{\epsilon} \non \\
  =&  \,\frac{1}{2 \pi i} \int_{\Gamma}
        \frac{\eta^{\mu \epsilon} d_{\sigma}(\zeta)}{\zeta} \,
        \big [\partial_{\alpha} (c^{~\beta}_{\epsilon~\gamma})\,
         \partial_{\mu} x^{\sigma}(-\zeta)
          + c^{~\beta}_{\mu~\gamma} \partial_{\alpha} \partial_{\epsilon}
          x^{\sigma} (-\zeta)\big ]\, d \zeta \\
     +& \frac{1}{2 \pi i}
       \int_{\Gamma} \frac{ b_{ \sigma \lambda
      }(\zeta)}{\zeta}\,\bigg [ \eta^{\mu \epsilon}
        c^{~\beta}_{\mu~\gamma}
        \partial_{\epsilon} x^{\lambda} (-\zeta) \partial_{\alpha}
        x^{\sigma} (\zeta)
        +  \eta^{\beta \epsilon} c^{~\nu}_{\alpha~\gamma}
          \partial_{\nu} x^{\lambda} (-\zeta) \partial_{\epsilon}
           x^{\sigma} (\zeta)\non \\
     & +  \eta^{\xi \epsilon} c^{~\beta}_{\alpha~\xi}
        \partial_{\epsilon} x^{\lambda} (-\zeta) \partial_{\gamma} x
        ^{\sigma}(\zeta)
       +  \eta^{\mu \epsilon} \partial_{\alpha} (c^{~\beta}_{\epsilon~\gamma})\,
         \partial_{\mu} x^{\lambda}(-\zeta) x^{\sigma}(\zeta)
       + \eta^{\mu \epsilon} c^{~\beta}_{\mu~\gamma}
       \partial_{\alpha} \partial_{\epsilon} x^{\lambda}(-\zeta)
       x^{\sigma} (\zeta)\bigg ] \,  d \zeta. \non
\end{align}
Taking into account (\ref{cf}), we see  that (\ref{cvari}) leads
to the expression (\ref{fvari}) for $\delta F$.

Let us assume now that the first element of the basis of $\cH$
is the unit element  and $\eta _{\alpha \beta} =
c_{1\alpha \beta}$. We would like to require that these
assumptions are fulfilled also for the deformed solution of
WDVV equations. Then we should have
\begin{align}
\label{var_eta3}
  \delta (\delta^{\beta}_{~\alpha})
  = \,  \delta c^{~\beta}_{\alpha ~ 1} (\tau)
  = \frac{1}{2 \pi i} \int_{\Gamma}
        \frac{\, \eta^{\beta \epsilon} \big [d_{\sigma}(\zeta)
        - \zeta b_{1 \sigma}(\zeta)\big ]}{\zeta} \,
        \partial_{\alpha} S^{\sigma}_{~\epsilon} (-\zeta) \, d \zeta
        =0.
\end{align}
This is true if we have the relation (\ref{dandb}) between the functions
$d(z)$ and $b(z)$.
We have seen in the introduction how to change the expressions for
$\delta F$ and $\delta x^{\alpha}$ if we require the existence of
unit element. We came again to the same prescription.

The monomials
$$b_{\alpha \beta,m} z^{-m}$$
with coefficients obeying
\begin{align}
  b_{\alpha \beta, m} = (-1)^{m+1} b_{\beta \alpha, m},
\end{align}
can be considered as (topological) generators of the Lie algebra
$\cB$. Let us write down corresponding variations of $F$ and of
coefficients $h_n$ that appear in the decomposition of
$x^{\alpha}(\tau,z)$  in a series $x^{\alpha}(\tau,z) =
\sum_{n=0}^{\infty} h^{\alpha,n} z^{-n}$. It follows from
(\ref{fvari}), (\ref{xvari}) and (\ref{dandb}) that
\begin{align}
\label{result_F}
  \delta F (\tau)
=& \,b_{\alpha 1,2}\, \tau^{\alpha}+ b_{\alpha 1,1}\, h^{\alpha,1}
   - \frac{1}{2} \, b_{\alpha \beta, 1}\, \tau^{\alpha}\tau^{\beta}
   + b_{\alpha 1,0}\, h^{\alpha, 2}
   + b_{\alpha \beta, 0}\, \tau^{\alpha} h^{\beta,1}\non \\
  &+  \sum_{m<0} \Bigg \{b_{\alpha 1,m}\, h^{\alpha, -m+2}
   + b_{\alpha \beta, m}\,
   \Big[ \,  (-1)^{m} \tau^{\alpha} h^{\beta, -m+1}
   + \frac{1}{2}\sum^{-1}_{n=m} (-1)^{m+n} h^{\alpha, -n}
    h^{\beta, -m+n+1} \Big ]\Bigg \} \\
\label{result_x}
   \delta x^{\alpha}(\tau,z)
  =& \, \eta^{\rho \epsilon} \sum_{k \geqslant 0} \partial_{\rho} h^{\alpha, k} z^{-k} \,
    \sum_{i \geqslant 0}
   \sum_{l=0}^{i}
  \partial_{\epsilon} h^{\sigma,l} \,(-1)^{l}
  \big (\,   b_{1 \sigma, -i+1}\,
   +\sum_{m \leqslant -i}   b_{\lambda \sigma, m}\,
  h^{\lambda,-m-i}  \big )\, z^{i-l}  \non \\
  & - \sum_{m =-\infty}^{\infty} \, \eta^{\alpha \sigma }\big [ \, z\, b_{1 \sigma, m}\,
   + b_{\lambda \sigma, m}\,
     x^{\lambda}(\tau, z) \big ]  z^{-m} \\
  \delta h^{\alpha, n}
  =&\,  \eta^{\rho \epsilon} \sum_{ 0 \leqslant l \leqslant i \leqslant \infty}\,
       \partial_{\rho} h^{\alpha, i+n -l}
      \partial_{\epsilon} h^{\sigma, l} (-1)^l \, b_{1 \sigma,
      -i+1} \non \\
  &   +  \eta^{\rho \epsilon} \sum_{0 \leqslant l \leqslant i \leqslant -m \leqslant \infty}
      \, \partial_{\rho} h^{\alpha, i+n -l}
      \partial_{\epsilon} h^{\sigma, l} (-1)^l \,
      b_{\lambda \sigma, m} h^{\lambda, -m -i} \non \\
  &- \eta^{\alpha \sigma} \, \big [\, b_{1 \sigma, n+1} + \sum_{m \leq n} b_{\lambda \sigma, m}\, h^{\lambda, n-m}\, \big ]
\end{align}
In particular, for $m=1$ we obtain
\begin{align}
  \delta F (\tau)
 =& \,b_{\alpha 1,1}\, h^{\alpha,1}
   - \frac{1}{2} \, b_{\alpha \beta, 1}\, \tau^{\alpha}\tau^{\beta},
  \non \\
  \delta c_{\alpha \beta \gamma} (\tau)
 =& \, \eta^{\lambda \epsilon} \,b_{\lambda 1,1}\,
    \partial_{\alpha}c_{\epsilon \beta \gamma} \, , \\
\delta \tau^{\alpha}= \delta h^{\alpha,0} =& \, 0, \non \\
\delta h^{\alpha, n}=
  &\,  \eta^{\rho \sigma} b_{1 \sigma, 1}\, \partial_{\rho} h^{\alpha, n}
  - \eta^{\alpha \sigma}  b_{\lambda \sigma, 1}\, h^{\lambda, n-1}. ~~~~~ n\geqslant 1 \non \\
\end{align}
For $m=2$
\begin{align}
\delta F (\tau)=&\, b_{\alpha 1,2}\, \tau^{\alpha}, \non \\
\delta c_{\alpha \beta \gamma} (\tau)=&\, 0, \\
\delta \tau^{\alpha}= \delta h^{\alpha,0} =& \, 0, \non \\
\delta h^{\alpha, 1}=
  &- \eta^{\alpha \sigma} b_{1 \sigma, 2}, \non \\
\delta h^{\alpha, n}=
  &- \eta^{\alpha \sigma}  b_{\lambda \sigma, 2}\, h^{\lambda, n-2}. ~~~~~ n\geqslant 2 \non \\
\end{align}
For $m>2$ we obtain
\begin{align}
\delta F (\tau)=& \, 0,\non \\
\delta c_{\alpha \beta \gamma} (\tau)=&\, 0,\\
\delta h^{\alpha,n} =&\, 0, ~~~~~~~~~~~~~~~~~~~~~~~~~~ 
  0\leqslant n \leqslant m-2 \non \\
\delta h^{\alpha, m-1}=
  &- \eta^{\alpha \sigma} \, b_{1 \sigma, m}, \non \\
\delta h^{\alpha, n}=
  &- \eta^{\alpha \sigma} \, b_{\lambda \sigma, m}\, h^{\lambda, n-m}. ~~~~~ n\geqslant m \non \\
\end{align}

\section{Comparison with Givental's approach}

\label{Giv}

\setcounter{equation}{0}

Givental's construction of twisted loop group acting on the space of
Frobenius manifolds is based on another geometric interpretation of the
notion of Frobenius structure.  For every calibrated solution of
WDVV equation one can construct a functional $\cF$ that depends on
infinite number of parameters $q^0, ...,q^n, ...$, where $q^k \in \cH$.
(The functional $\cF$ can be interpreted as free energy depending on
primary fields and descendants. The construction of $\cF$ in
terms of free energy $F$, defined on "small phase space" $\cH$, and calibration is given in [5],
[1]). More precisely, one should
consider
$\cF$ as a formal power series with respect to its arguments; it is
important to emphasize that parameters $q^k$ used by Givental are obtained
from standard parameters $\tau^0, ..., \tau^n, ...$ in the expression of free
energy by means of a shift in $\tau^1$.  Notice , that in Givental's paper
it is assumed that there exists a unit element and $\eta _{\alpha \beta}=c_{1\alpha \beta}$.
(This is a standard assumption relaxed in present paper mostly to simplify the exposition.)

Let us introduce a space of $\cH$-valued Laurent polynomials
$a(z)$ equipped with symplectic form
\begin{align}
\label{form}
\omega (a,b) = res (a(z),b(-z))=(2\pi i)^{-1} \int_{\Gamma}
(a(z),b(-z)) dz=(2\pi i)^{-1} \int_{\Gamma} a^{\alpha}(z) \eta
_{\alpha \beta} b^{\beta}(-z) dz
\end{align}.

This space can be identified with $T^{\ast}L$ where $L$ is
a Lagrangian subspace consisting of polynomials $q^0+q^1 z+ q^2 z^2+...$.
The functional $\cF$ can be considered as a function on $L$; we
construct a Lagrangian submanifold $\cL$ corresponding to $\cF$ by means
of standard formula
\begin{align}
p_i=\partial \cF/ \partial q^i.
\end{align}
One can prove [2], that $\cL$ is a Lagrangian cone with the vertex at
the origin and that every tangent space $T_x \cL$ to $\cL$ is tangent
to $\cL$ exactly along $zT_x \cL$. (More precisely, if $K$ is a tangent
space to $\cL$ then its intersection with $\cL$ consists of points of
the form $zk$ where $k \in K$.) Conversely, every Lagrangian cone
obeying these conditions corresponds to a calibrated solution of WDVV
equations. Givental defines twisted loop group as a group of matrix
Laurent series satisfying
\begin{align}
\label{M}
  M(z) M^*(-z)&=1.
\end{align}
It is easy to check that elements of this group can be interpreted as
linear symplectic transformations commuting with multiplication by $z$.
Lagrangian cones corresponding to calibrated solutions of WDVV equations
are defined in terms of symplectic geometry and multiplication by $z$.
This means that an element of twisted loop group transforms such a cone
into
another cone of the same kind and a calibrated solution of WDVV equations
into another calibrated solution.(This is a rigorous statement if one
works at the level of Lie algebras; to define the action of an element
of twisted loop group one should be more precise with the definition of
free energy in terms of formal series.)
Using the construction of $\cF$ in terms
of $F$ and calibration one can calculate the action of Lie algebra of
twisted loop group on calibrated solutions of WDVV equations; one gets
formulas that agree with (3.15),(3.16). One should
emphasize, however, that Givental works with formal series instead of
functions.

Let us discuss the relation between Givental's geometric picture and
the setup of the present paper.  We will work with symplectic vector
space $\cal C$ of $\cal H$-valued holomorphic functions on
$\bC^{\times}$ having a pole or a removable singularity at infinity;
the symplectic form on $\cal C$ is defined by the formula (\ref{form}).

We start with a calibrated solution of WDVV equations and define a
subset $\cal L$ of $\cal C$ as a set of all functions of the form $z
p^{\lambda} \frac{\partial x(\tau, z)}{\partial \tau^{\lambda}}$, where $\tau
\in \cal T$, $x(\tau,z)=(x^1(\tau,z),..., x^n(\tau,z))$ is a
calibration and $p^{\lambda}(z)$ is a polynomial.  One can say that
$\cal L$ is a union of vector spaces ${\cal L}_{\tau} = \big \{
z p^{\lambda} \frac{\partial x(\tau, z)}{\partial \tau^{\lambda}} \vert
p^{\lambda} \in \bC[z] \big \}$ labelled by $\tau \in \cal T$.  It is
easy to check that $\cal L$ is an isotropic submanifold of $\cal C$.
Indeed, the tangent space $L_{f} = {\cal T}_f \cal L$ to $\cal L$ at
the point $f= z p^{\lambda}(z) \frac{\partial x(\tau, z)}{\partial
\tau^{\lambda}}\neq 0$ is spanned by $\cal L_{\tau}$ and $\partial
f/\partial \tau^1, ... \partial f / \partial \tau^n$.  Using the
relation
\begin{align}
\label{rel_1}
  \frac{\partial f}{\partial \tau^{\sigma}}
  = z p^{\lambda}(z) \frac{\partial^2
    x(\tau, z)}{\partial \tau^{\lambda} \partial \tau^{\sigma}}
  = p^{\lambda}(z) c^{~ \gamma}_{\lambda ~ \sigma}
    \frac{\partial x(\tau, z)}{\partial \tau^{\gamma}}
\end{align}
and (\ref{normalization1}) we find that the form (\ref{form}) vanishes
on all tangent spaces.

Let us suppose now that $\eta_{\alpha \beta}= c_{1 \alpha \beta}$ and
that the function $x^{\alpha}$ obeys the normalization condition
(\ref{normalization2}).  Introducing the notation $J^{\mu}(\tau,z) =
x^{\mu}(\tau, z)+z$, we conclude from (\ref{normalization2}) that
the point
\begin{align}
  J(\tau, z) = z \partial_1 J(\tau,z)
\end{align}
belongs to $\cal L_{\tau}$(it can be represented as $f=z
p^{\lambda}(z) \frac{\partial x(\tau, z)}{\partial \tau^{\lambda}}$
with $p(z)= (1,0,...,0)$).  Applying  (\ref{rel_1}) we see that the
tangent space $L_f = T_f \cal L$ at this point is spanned by $\cal
L_{\tau}$ and $\partial f/\partial \tau^{\sigma} = \partial x
/\partial \tau^{\sigma}$, in other words, it consists of vectors of
the form $p^{\lambda}(z) \frac{\partial x}{\partial \tau^{\lambda}}$, where
$p^{\lambda}$ is an arbitrary polynomial.  Using (1.18) we obtain that
this tangent space is Lagrangian.  The isotropic submanifold $\cal L$
is a union of vector spaces, hence it is always a cone. The 
subset of $\cal L$ consisting of points  where the
matrix $p^{\lambda}(0) c^{~ \gamma}_{\lambda ~ \sigma}$ is
nondegenerate can be regarded as Lagrangian cone. Notice that this subset can be empty if  we do not assume the existence of unit element ; it is
non-empty if for at least one element of  the algebra with structure
constants $ c^{~ \gamma}_{\lambda ~ \sigma}$ the operator of multiplication
by this element is invertible. (We use the fact that an isotropic subspace of  $ T^{\ast}L$
is Lagrangian if its projection on $L$ is surjective.)

Let us describe the way to obtain calibrated solutions from
Lagrangian  cones following the considerations applied in [2] to the
case when there exists a unit element. Let us consider  a
Lagrangian  cone  $\cL \subset T^{\ast}L$ with the vertex at the
origin assuming   that every tangent space $T_x \cL$ to $\cL$ is
tangent to $\cL$ exactly along $zT_x \cL$. Let us assume that for
every $\tau$ the cone $\cL$ contains a point $J(z,\tau)=cz+\tau +
r(z,\tau)$ where $r(\infty,\tau)=0$ . Then the derivatives $\partial J
/\partial \tau ^{\alpha}$ form a basis in  $\Lambda /z \Lambda$
where $\Lambda$ stands for one of these tangent spaces.  They can
be regarded also as free generators of $\Lambda$ considered as
$\bC [z]$-module.  Using the fact that
 $z\partial J/ \partial \tau \in  z\Lambda \subset \cL$ we obtain that the second derivatives
$z\partial ^2 J /\partial \tau ^{\alpha } \partial \tau ^{\beta}$ are in
$\Lambda$ and therefore can be represented as linear combinations
of  $\partial J /\partial \tau ^{\alpha}$  with coefficients that
are polynomials with respect to $z$. From the other side these
second derivatives have a removable singularity at infinity, hence
the coefficients do not  depend on $z$. These coefficients
$c^{~\gamma}_{\alpha ~\beta}$ specify a family of torsion-free
flat connections  by the formula (1.7). (The connections are  flat
because the equation (1.9) has a non-degenerate solution $S(z,
\tau)= \partial J / \partial \tau$.)

The connections we constructed together with $J(z, \tau)$ specify
a calibrated solution to the WDVV equations.  To finish the proof of
this statement we should check (2.1). However, one can verify that
(2.1) follows from (1.18). ( Both of these conditions can be
interpreted geometrically as compatibility of non-degenerate
linear pairing between tangent spaces to  affine spaces $T_z$ and $T_{-z}$  with the connection
 $\Gamma_{\alpha \beta}^{\gamma}
(z)=z^{-1} c_{\alpha ~\beta}^{~\gamma}$.)  We used (1.18) to prove
that the cone constructed by means of a calibrated solution to
WDVV equations is isotropic; one can use the same arguments in
opposite direction: to show that the family of connections
obtained from a  Lagrangian cone obeys (1.18).

Acknowledgments. We are indebted to A. Givental for
interesting discussions. The work of A. Sch. was
partially supported by NSF grant DMS-0204927.

\appendix
\section{Appendix. Calculations in Givental's approach}
\label{sec:results}
\setcounter{equation}{0}

Denote by $\cL$ the Lagrangian cone corresponding  to free energy $\cF$:
\begin{align}
  \cL = \{ (\bp, \bq) \in T^* L: \bp = d_{\bq} \cF \}.
\end{align}

Introduce Darboux coordinates $\{p^{\alpha,n}, q^{\alpha,n}\}$,
$n=0,1,...$ and $\alpha = 1,..., N$, with
\begin{align}
\label{generating}
  p_{\alpha,n} = \frac{\partial \cF}{\partial q^{\alpha,n}}
\end{align}
on the Lagrangian cone $\cL$. Here
$q^n$ are elements of a finite-dimensional vector space denoted by
$\cH$. We can explicitly choose a basis in $\cH$ and  work in
components $q^{\alpha,i}$. Note that $q$ corresponds to the unshifted
$T$ in Dubrovin's paper [1].

Let us consider an element of Lie algebra of the twisted loop group.
As every linear infinitesimal symplectic transformation it corresponds
to a quadratic Hamiltonian  $H$ :
\begin{align}
\label{eom}
  \delta q^{\alpha,n} &= \frac{\partial H}{\partial p_{\, \alpha,n}}, \non \\
  \delta p_{\alpha,n} &= -\frac{\partial H}{\partial q^{\, \alpha,n}}.
\end{align}
The variation of the generating function $\cF$
is given by the formula
\begin{align}
\label{def_vari_F}
  \delta \cF  \, = - H \big ( \frac{\partial \cF}{\partial
q^{\alpha, n}}, q^{\beta, n'}
  \big ) .
\end{align}
We are interested in the variation of $\cF$ restricted to the small
phase space(SPS)
where
\begin{align}
\label{sps}
   q^{\alpha, 0} =\tau^{\alpha}, ~~~~~~ q^{1,1} =-1,  ~~~~~~q^{\alpha,
i} =0 , ~~i >1.
\end{align}

It is convenient to represent an element of $T^* L $ as a Laurent  series
with coefficients $q^n$ with $n \in {\mathbb Z}$. Then
we can write general quadratic hamiltonian in the  form
\begin{align}
  H = \frac{1}{2} \, b_{\alpha \beta, mn}\, q^{\alpha, m} q^{\beta, n},
\end{align}
where summation over indices is implicit and
\begin{align}
\label{b_symmetry}
   b_{\alpha \beta, mn} =  b_{\beta \alpha, nm}.
\end{align}
If $H$ corresponds to an element of Lie algebra of twisted loop group it should
be invariant under the multiplication by $z$. This means that it has the form
\begin{align}
\label{generator}
  H = \frac{1}{2} \, \sum_{m} b_{\alpha \beta, m}\,
      \sum_{n}  (-1)^{n} q^{\alpha, n} q^{\beta, m-n-1},
\end{align}
where
\begin{align}
  b_{\alpha \beta, m} = (-1)^{m-1} b_{\beta \alpha, m} .
\end{align}

From (\ref{generating}) and the Poisson bracket (4.1)
\begin{align}
   \{q^{\alpha,m}, q^{\beta,n} \} = (-1)^{n}\eta^{\alpha \beta} \delta^{m+n+1},
\end{align}

we have that on $\cL$
\begin{align}
\label{qp}
    q^{\alpha, n} = (-1)^{n} \eta^{\alpha \beta}
            \frac{\partial \cF}{\partial q^{\beta, -n-1}}
\end{align}
 for $n<0$.
The free energy $\cF$ is identified with the logarithm of the
$\tau$-function, which can be written as
\begin{align}
\label{tau_fun}
  \log \tau_0 (q) = \frac{1}{2} \, \underset{z^{-1}=0}{\res} \;
                    \underset{w^{-1}=0}{\res} \, \sum_{r,s}
                    z^{r+1} w^{s+1} q^{\lambda, r} q^{\mu, s}
                    V_{\lambda \mu},
\end{align}
where
\begin{align}
  V_{\alpha \beta}(t; z,w) = \sum_{p,q} V_{(\alpha,p)(\beta, q)}
  z^{-p} w^{-q}.
\end{align}
In particular, we will need the following coefficients calculated in [1]
\begin{align}
  V_{(\alpha, 0)(\beta,0)} &= F_{\alpha \beta}, \non \\
  V_{(\alpha, p)(\beta,0)} &= \partial_{\beta} h_{\alpha, p+1}, \non \\
  V_{(\alpha, p)(1,1)} &= (\tau^{\lambda}
      \partial_{\lambda} -1) \,h_{\alpha, p+1} , \non \\
  V_{(\alpha, 0)(1,1)} &= F_{\alpha \lambda} \tau^{\lambda} -
                           F_{\alpha}, \non \\
  V_{(1, 1)(1,1)} &=  F_{\lambda \mu} \tau^{\lambda} \tau^{\mu} - 2
                      F_{\lambda} \tau^{\lambda} + 2F,
\end{align}
where $F$ stands for the free energy resticted to the small phase space (SPS):
\begin{align}
  F = \cF\bvert_{\sps}.
\end{align}
Explicitly we have from eqn.(\ref{tau_fun}) and (\ref{sps})
\begin{align}
  \partial_{q^{\alpha, p}} (\log \tau_0) \bvert_{\sps}
  \; =&\;  \underset{z^{-1}=0}{\res} \; \underset{w^{-1}=0}{\res} \,
    z^{p+1}  \sum_{s} w^{s+1}\, q^{\mu, s} \,
    V_{\alpha \mu} \bvert_{\sps} \non \\
    &+ \frac{1}{2} \, \underset{z^{-1}=0}{\res} \; \underset{w^{-1}=0}{\res} \,
       \sum_{r,s} z^{r+1} w^{s+1} \,q^{\lambda, r} \,q^{\mu, s} \,
       \frac{\partial V_{\lambda \mu}}{\partial \tau^{\sigma}} \,
       \frac{\partial \tau^{\sigma}}{\partial q^{\alpha,p}}\bvert_{\sps}
       \non \\
   =&\;  h_{\alpha, p+1}.
\end{align}
Together with (\ref{qp}) we have
\begin{align}
   q^{\alpha, n} = (-1)^{n} h^{\alpha, -n}, ~~~~ n<0.
\end{align}
Put this expression into (\ref{def_vari_F}), we now obtain the
variation of the free energy on the small phase space
\begin{align}
\label{F_variation}
  \delta \cF \bvert_{\sps}
 = & - \frac{1}{2} \, \Big [\, b_{11,3}(-1) q^{1,1} q^{1,1}
    + \sum_{m \leq 2} b_{\alpha 1, m} (-1)^{m} q^{\alpha,m-2} q^{1,1}
   + \sum_{m \leq 2} b_{1 \beta, m} (-1) q^{1,1} q^{\beta,m-2} \non \\
   & +\sum_{m \leq 1} b_{\alpha \beta, m}
    \sum^{0}_{n = m-1} (-1)^n q^{\alpha, n} q^{\beta, m-n-1} \Big ]  \non \\
=& b_{11,3}/2
   +   b_{\alpha 1,2}\, \tau^{\alpha} +  b_{\alpha 1,1}\, h^{\alpha,1}
   - \frac{1}{2} \,  b_{\alpha \beta, 1}\, \tau^{\alpha}\tau^{\beta}
   + b_{\alpha 1,0}\, h^{\alpha, 2}
   + b_{\alpha \beta, 0}\, \tau^{\alpha} h^{\beta,1}\non \\
  &+  \sum_{m<0} \Bigg \{ b_{\alpha 1,m}\, h^{\alpha, -m+2} +
b_{\alpha \beta, m}\,
   \Big[  (-1)^{m} \tau^{\alpha} h^{\beta, -m+1}
   + \frac{1}{2} \, \sum^{-1}_{n=m} (-1)^{m+n} h^{\alpha, -n}
    h^{\beta, -m+n+1} \Big ]\Bigg \} .
\end{align}
The variation of the structure constant is given by the third derivative
\begin{align}
\label{c_variation}
 &\delta c_{\alpha \beta \gamma}
 = \partial_{\alpha} \partial_{\beta} \partial_{\gamma}
   \delta \cF \bvert_{\sps} \non \\
 =& \partial_{\alpha} \partial_{\beta} \partial_{\gamma}
    \Big [\, b_{\mu 1,1}\, h^{\mu,1}
    + b_{\mu 1,0}\, h^{\mu, 2}
    + b_{\mu \nu, 0}\, \tau^{\mu} h^{\nu,1} \Big ]\non \\
   +& \sum_{m<0}\partial_{\alpha} \partial_{\beta} \partial_{\gamma}
     \Bigg \{\, b_{\mu 1,m}\, h^{\mu, -m+2}  + b_{\mu \nu, m}\,
   \Big[ \,  (-1)^{m} \tau^{\mu} h^{\nu, -m+1}
   + \frac{1}{2}\,\sum^{-1}_{n=m} (-1)^{m+n} h^{\mu, -n}
    h^{\nu, -m+n+1} \Big ]\Bigg \}.
\end{align}

The equations (\ref{F_variation}) agrees with equations
(\ref{result_F}), up to a trivial constant term.



\begin{thebibliography}{99}

\bibitem{dubrovin} B. Dubrovin, Nucl. Phys. {\bf B}379, 627 (1992)

\bibitem{givental1} A. Givental, math.AG/0305409

\bibitem{bar1} S. Barannikov, math.AG/0006193

\bibitem{leur} J. van de Leur, nlin.SI/0004021

\bibitem{DW} R.Dijkgraaf, E. Witten,Nucl.Phys, B342,486 (1990)

\end{thebibliography}
\end{document}